%% file: main.tex
\documentclass[12pt,a4paper]{article}

\usepackage{ifthen} 
\newboolean{pdflatex}
\setboolean{pdflatex}{false} 

\newboolean{articletitles}
\setboolean{articletitles}{true} 

\newboolean{uprightparticles}
\setboolean{uprightparticles}{false} 

\input{preamble}
\input{shortcuts}

\begin{document}

\renewcommand{\thefootnote}{\fnsymbol{footnote}}
\setcounter{footnote}{1}

\input{title-LHCb-PAPER}


\renewcommand{\thefootnote}{\arabic{footnote}}
\setcounter{footnote}{0}


\pagestyle{plain} 
\setcounter{page}{1}
\pagenumbering{arabic}

\input{introduction}
\input{selection}
\input{background}
\input{extraction}

\input{CPAsymmetry}
\input{conclusions}
\input{aknowledgments}

\addcontentsline{toc}{section}{References}
\bibliographystyle{LHCb}
\bibliography{references}
\end{document}

%% file: preamble.tex
\usepackage{microtype}
\usepackage{lineno}
\usepackage{xspace} 

\usepackage{graphicx}  
\usepackage{color}
\usepackage{colortbl}

\usepackage{amsmath} 
\usepackage{amssymb}
\usepackage{amsfonts}
\usepackage{upgreek} 

\newcommand\Tstrut {\rule{0pt}{2.6ex}}
\newcommand\TTstrut{\rule{0pt}{3.2ex}}
\newcommand\Bstrut {\rule[-1.2ex]{0pt}{0pt}}
\newcommand\BBstrut{\rule[-1.8ex]{0pt}{0pt}}

\newcommand*\patchAmsMathEnvironmentForLineno[1]{%
\expandafter\let\csname old#1\expandafter\endcsname\csname #1\endcsname
\expandafter\let\csname oldend#1\expandafter\endcsname\csname
end#1\endcsname
 \renewenvironment{#1}%
   {\linenomath\csname old#1\endcsname}%
   {\csname oldend#1\endcsname\endlinenomath}%
}
\newcommand*\patchBothAmsMathEnvironmentsForLineno[1]{%
  \patchAmsMathEnvironmentForLineno{#1}%
  \patchAmsMathEnvironmentForLineno{#1*}%
}
\AtBeginDocument{%
\patchBothAmsMathEnvironmentsForLineno{equation}%
\patchBothAmsMathEnvironmentsForLineno{align}%
\patchBothAmsMathEnvironmentsForLineno{flalign}%
\patchBothAmsMathEnvironmentsForLineno{alignat}%
\patchBothAmsMathEnvironmentsForLineno{gather}%
\patchBothAmsMathEnvironmentsForLineno{multline}%
}

\usepackage{hyperref}    
\usepackage[all]{hypcap} 

\input{lhcb-symbols-def} 

\usepackage{cite} 
\usepackage{mciteplus}

%% file: lhcb-symbols-def.tex
\def\lhcb {\mbox{LHCb}\xspace}
\def\ux85 {\mbox{UX85}\xspace}

\def\lhc    {\mbox{LHC}\xspace}

\ifthenelse{\boolean{uprightparticles}}%
{
 
 \def\Pgamma      {\ensuremath{\upgamma}\xspace}

 \def\Ppi         {\ensuremath{\uppi}\xspace}

 \def\Pphi        {\ensuremath{\upphi}\xspace}

 \def\Ppsi        {\ensuremath{\uppsi}\xspace}

 \def\PDelta      {\ensuremath{\Delta}\xspace}                 
 \def\PXi      {\ensuremath{\Xi}\xspace}                 
 \def\PLambda      {\ensuremath{\Lambda}\xspace}                 
 \def\PSigma      {\ensuremath{\Sigma}\xspace}                 
 \def\POmega      {\ensuremath{\Omega}\xspace}                 
 \def\PUpsilon      {\ensuremath{\Upsilon}\xspace}                 
 
 \def\PB      {\ensuremath{\mathrm{B}}\xspace}                 
                  
 \def\PD      {\ensuremath{\mathrm{D}}\xspace}

 \def\PJ      {\ensuremath{\mathrm{J}}\xspace}                 
 \def\PK      {\ensuremath{\mathrm{K}}\xspace}

 \def\PW      {\ensuremath{\mathrm{W}}\xspace}

 \def\Pb      {\ensuremath{\mathrm{b}}\xspace}                 
 \def\Pc      {\ensuremath{\mathrm{c}}\xspace}

 \def\Pi      {\ensuremath{\mathrm{i}}\xspace}

 \def\Pp      {\ensuremath{\mathrm{p}}\xspace}

 \def\Ps      {\ensuremath{\mathrm{s}}\xspace}

}
{
 
 \def\Pgamma      {\ensuremath{\gamma}\xspace}

 \def\Ppi         {\ensuremath{\pi}\xspace}

 \def\Pphi        {\ensuremath{\phi}\xspace}

 \def\Ppsi        {\ensuremath{\psi}\xspace}                 
                  
 \mathchardef\PDelta="7101
 \mathchardef\PXi="7104
 \mathchardef\PLambda="7103
 \mathchardef\PSigma="7106
 \mathchardef\POmega="710A
 \mathchardef\PUpsilon="7107
                  
 \def\PB      {\ensuremath{B}\xspace}                 
                  
 \def\PD      {\ensuremath{D}\xspace}

 \def\PJ      {\ensuremath{J}\xspace}                 
 \def\PK      {\ensuremath{K}\xspace}

 \def\PW      {\ensuremath{W}\xspace}

 \def\Pb      {\ensuremath{b}\xspace}                 
 \def\Pc      {\ensuremath{c}\xspace}

 \def\Pi      {\ensuremath{i}\xspace}

 \def\Pp      {\ensuremath{p}\xspace}

 \def\Ps      {\ensuremath{s}\xspace}

}

\def\g      {\ensuremath{\Pgamma}\xspace}

\def\W      {\ensuremath{\PW}\xspace}

\def\squark    {\ensuremath{\Ps}\xspace}

\def\cquark    {\ensuremath{\Pc}\xspace}

\def\bquark    {\ensuremath{\Pb}\xspace}

\def\pion  {\ensuremath{\Ppi}\xspace}
\def\piz   {\ensuremath{\pion^0}\xspace}

\def\pip   {\ensuremath{\pion^+}\xspace}
\def\pim   {\ensuremath{\pion^-}\xspace}

\def\pipm  {\ensuremath{\pion^\pm}\xspace}

\def\kaon  {\ensuremath{\PK}\xspace}
  \def\Kbar  {\kern 0.2em\overline{\kern -0.2em \PK}{}\xspace}

\def\Kz    {\ensuremath{\kaon^0}\xspace}
\def\Kzb   {\ensuremath{\Kbar^0}\xspace}
\def\KzKzb {\ensuremath{\Kz \kern -0.16em \Kzb}\xspace}
\def\Kp    {\ensuremath{\kaon^+}\xspace}
\def\Km    {\ensuremath{\kaon^-}\xspace}

\def\KpKm  {\ensuremath{\Kp \kern -0.16em \Km}\xspace}

\def\Kstarz  {\ensuremath{\kaon^{*0}}\xspace}
\def\Kstarzb {\ensuremath{\Kbar^{*0}}\xspace}

  \def\Dbar    {\kern 0.2em\overline{\kern -0.2em \PD}{}\xspace}
\def\D       {\ensuremath{\PD}\xspace}

\def\Dz      {\ensuremath{\D^0}\xspace}
\def\Dzb     {\ensuremath{\Dbar^0}\xspace}
\def\DzDzb   {\ensuremath{\Dz {\kern -0.16em \Dzb}}\xspace}
\def\Dp      {\ensuremath{\D^+}\xspace}
\def\Dm      {\ensuremath{\D^-}\xspace}

\def\DpDm    {\ensuremath{\Dp {\kern -0.16em \Dm}}\xspace}

\def\Dstarpm {\ensuremath{\D^{*\pm}}\xspace}

\def\B       {\ensuremath{\PB}\xspace}
  \def\Bbar    {\kern 0.18em\overline{\kern -0.18em \PB}{}\xspace}

\def\Bu      {\ensuremath{\B^+}\xspace}

\def\Bp      {\ensuremath{\Bu}\xspace}

\def\Bd      {\ensuremath{\B^0}\xspace}
\def\Bs      {\ensuremath{\B^0_\squark}\xspace}
\def\Bsb     {\ensuremath{\Bbar^0_\squark}\xspace}
\def\Bdb     {\ensuremath{\Bbar^0}\xspace}

\def\jpsi     {\ensuremath{{\PJ\mskip -3mu/\mskip -2mu\Ppsi\mskip 2mu}}\xspace}

  \def\Y#1S{\ensuremath{\PUpsilon{(#1S)}}\xspace}

\def\proton      {\ensuremath{\Pp}\xspace}

\def\L {\ensuremath{\PLambda}\xspace}
\def\Lbar {\ensuremath{\kern 0.1em\overline{\kern -0.1em\PLambda}}\xspace}

\def\Lb      {\ensuremath{\L^0_\bquark}\xspace}

\def\BF         {{\ensuremath{\cal B}\xspace}}

\def\BR         {\BF}
\newcommand{\decay}[2]{\ensuremath{#1\!\to #2}\xspace}         

\def\to                 {\ensuremath{\rightarrow}\xspace}

\def\CP                {\ensuremath{C\!P}\xspace}

\def\BdToJPsiKst  {\decay{\Bd}{\jpsi\Kstarz}}

\def\BsPhiGam     {\decay{\Bs}{\phi \g}}
\def\BdKstGam     {\decay{\Bd}{\Kstarz \g}}

\def\AT#1     {\ensuremath{A_{\mathrm{T}}^{#1}}\xspace}           

\def\C#1      {\ensuremath{\mathcal{C}_{#1}}\xspace}                       
\def\Cp#1     {\ensuremath{\mathcal{C}_{#1}^{'}}\xspace}                    
\def\Ceff#1   {\ensuremath{\mathcal{C}_{#1}^{\mathrm{(eff)}}}\xspace}        
\def\Cpeff#1  {\ensuremath{\mathcal{C}_{#1}^{'\mathrm{(eff)}}}\xspace}       
\def\Ope#1    {\ensuremath{\mathcal{O}_{#1}}\xspace}                       
\def\Opep#1   {\ensuremath{\mathcal{O}_{#1}^{'}}\xspace}                    

\newcommand{\tev}{\ensuremath{\mathrm{\,Te\kern -0.1em V}}\xspace}
\newcommand{\gev}{\ensuremath{\mathrm{\,Ge\kern -0.1em V}}\xspace}
\newcommand{\mev}{\ensuremath{\mathrm{\,Me\kern -0.1em V}}\xspace}
\newcommand{\kev}{\ensuremath{\mathrm{\,ke\kern -0.1em V}}\xspace}
\newcommand{\ev}{\ensuremath{\mathrm{\,e\kern -0.1em V}}\xspace}
\newcommand{\gevc}{\ensuremath{{\mathrm{\,Ge\kern -0.1em V\!/}c}}\xspace}
\newcommand{\mevc}{\ensuremath{{\mathrm{\,Me\kern -0.1em V\!/}c}}\xspace}
\newcommand{\gevcc}{\ensuremath{{\mathrm{\,Ge\kern -0.1em V\!/}c^2}}\xspace}
\newcommand{\gevgevcccc}{\ensuremath{{\mathrm{\,Ge\kern -0.1em V^2\!/}c^4}}\xspace}
\newcommand{\mevcc}{\ensuremath{{\mathrm{\,Me\kern -0.1em V\!/}c^2}}\xspace}

\def\mum  {\ensuremath{\,\upmu\rm m}\xspace}

\def\invpb {\ensuremath{\mbox{\,pb}^{-1}}\xspace}

\def\invfb   {\ensuremath{\mbox{\,fb}^{-1}}\xspace}

\newcommand{\chisq}{\ensuremath{\chi^2}\xspace}

\def\gsim{{~\raise.15em\hbox{$>$}\kern-.85em
          \lower.35em\hbox{$\sim$}~}\xspace}
\def\lsim{{~\raise.15em\hbox{$<$}\kern-.85em
          \lower.35em\hbox{$\sim$}~}\xspace}

\def\sqs   {\ensuremath{\protect\sqrt{s}}\xspace}

\def\pt         {\mbox{$p_{\rm T}$}\xspace}
\def\et         {\mbox{$E_{\rm T}$}\xspace}

\def\mrad{\ensuremath{\rm \,mrad}\xspace}

\def\evtgen     {\mbox{\textsc{EvtGen}}\xspace}
\def\pythia     {\mbox{\textsc{Pythia}}\xspace}

\def\geant      {\mbox{\textsc{Geant4}}\xspace}

\def\photos     {\mbox{\textsc{Photos}}\xspace}

\def\tell1  {TELL1\xspace}
\def\ukl1   {UKL1\xspace}



%% file: shortcuts.tex
\def\photos{\mbox{\textsc{Photos}}\xspace} 

\ifthenelse{\boolean{uprightparticles}}%
{
 \def\Pphi      {\ensuremath{\upphi}\xspace}
}
{
 \def\Pphi      {\ensuremath{\phi}\xspace}
}

\def\pp    {\ensuremath{\proton\proton}\xspace}

\newcommand{\eq}[1]{Eq.~\ref{equation:#1}}

\newcommand{\fig}[1]{Fig.~\ref{figure:#1}}
\newcommand{\figf}[1]{Figure~\ref{figure:#1}}

\newcommand{\secr}[1]{Sect.~\ref{section:#1}}

\def\BRBdKstGam {\ensuremath{\BR(\decay{\Bd}{\Kstarz \g})}\xspace}
\def\BRBsPhiGam {\ensuremath{\BR(\decay{\Bs}{\phi \g})}\xspace}

%% file: title-LHCb-PAPER.tex
\begin{titlepage}
\pagenumbering{roman}

\vspace*{-1.5cm}
\centerline{\large EUROPEAN ORGANIZATION FOR NUCLEAR RESEARCH (CERN)}
\vspace*{1.5cm}
\hspace*{-0.5cm}
\begin{tabular*}{\linewidth}{lc@{\extracolsep{\fill}}r}
\ifthenelse{\boolean{pdflatex}}
{\vspace*{-2.7cm}\mbox{\!\!\!\includegraphics[width=.14\textwidth]{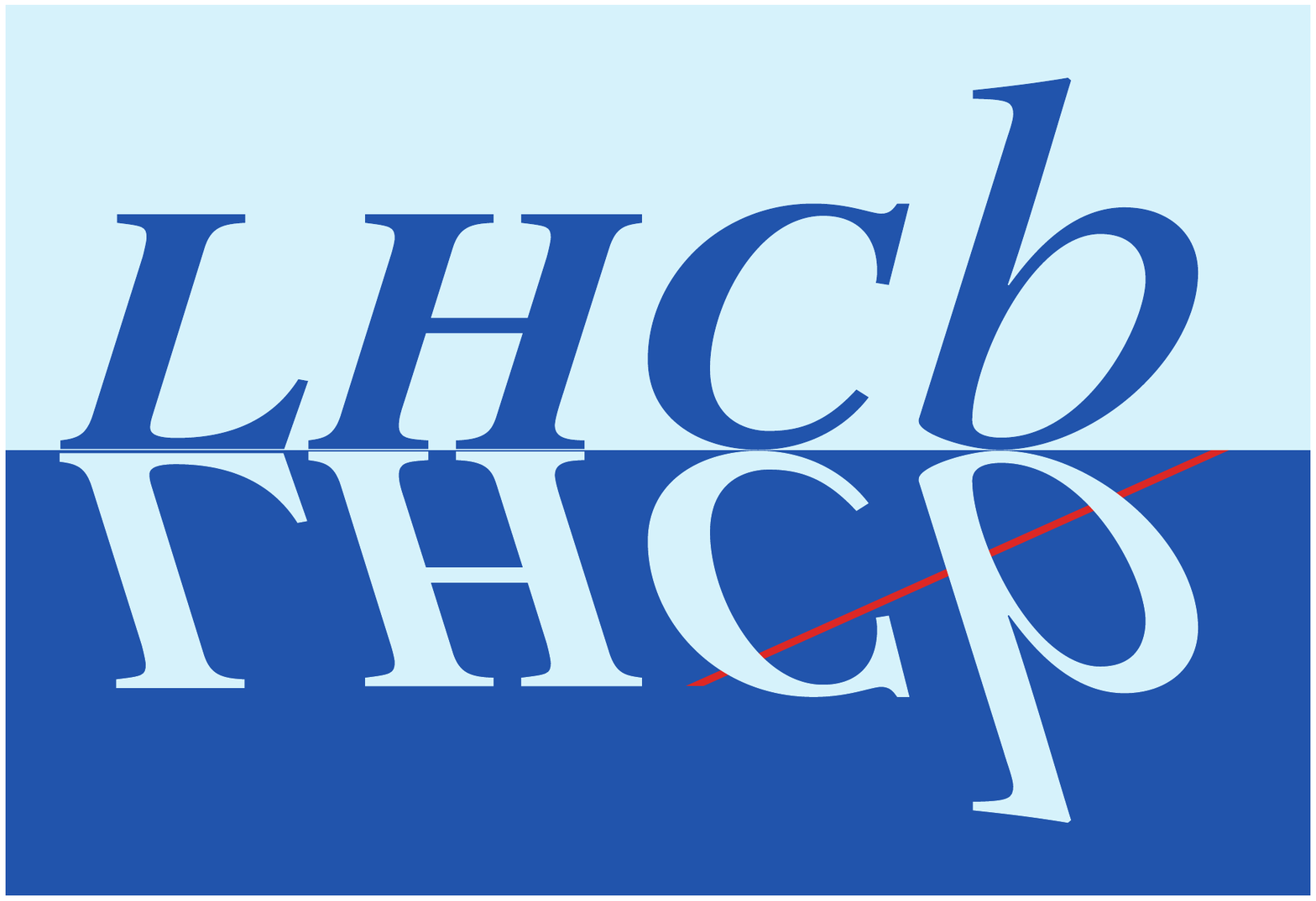}} & &}%
{\vspace*{-1.2cm}\mbox{\!\!\!\includegraphics[width=.12\textwidth]{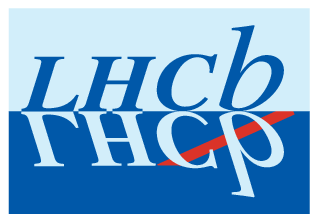}} & &}%
\\
 & & CERN-PH-EP-2012-247 \\  
 & & LHCb-PAPER-2012-019 \\  
 & & August 28, 2012 \\ 
 & & \\
\end{tabular*}

\vspace*{1.cm}

{\bf\boldmath\huge

\begin{center}
Measurement of the ratio of branching fractions \BRBdKstGam/\BRBsPhiGam and the direct \CP asymmetry in \BdKstGam 
\end{center}
}

\vspace*{2.cm}

\begin{center}
The LHCb collaboration\footnote{Authors are listed on the following pages.}
\end{center}

\vspace{\fill}

\begin{abstract}
\noindent
The ratio of branching fractions of the radiative \B decays \BdKstGam and \BsPhiGam has been measured using an integrated luminosity of 1.0\invfb of \pp collision data collected by the \lhcb experiment at a centre-of-mass energy of \sqs$=7$\tev. The value obtained is 
\begin{equation*}
\frac{\BRBdKstGam}{\BRBsPhiGam} = 1.23 \pm 0.06\mathrm{\,(stat.)} \pm 0.04\mathrm{\,(syst.)} \pm 0.10\,(f_s/f_d)\,,
\end{equation*}
where the first uncertainty is statistical, the second is the experimental systematic uncertainty and the third is associated with the ratio of fragmentation fractions $f_s/f_d$.
Using the world average value for \BRBdKstGam, the branching fraction \mbox{\BRBsPhiGam} is measured to be $(3.5\pm 0.4)\times10^{-5}$.
 
The direct \CP asymmetry in \BdKstGam decays has also been measured with the same data and found to be
\begin{equation*}
\mathcal{A}_{\CP}(\BdKstGam) = (0.8\pm1.7\mathrm{\,(stat.)}\pm0.9\mathrm{\,(syst.)})\%\,.
\end{equation*}

Both measurements are the most precise to date and are in agreement with the previous experimental results and theoretical expectations. 
\end{abstract}

\begin{center}
Submitted to Nuclear Physics B
\end{center}

\vspace{\fill}

\end{titlepage}


\newpage
\setcounter{page}{2}
\mbox{~}
\newpage

\input{LHCb_authorlist.tex}

\cleardoublepage

%% file: LHCb_authorlist.tex
\centerline{\large\bf LHCb collaboration}
\begin{flushleft}
\small
R.~Aaij$^{38}$, 
C.~Abellan~Beteta$^{33,n}$, 
A.~Adametz$^{11}$, 
B.~Adeva$^{34}$, 
M.~Adinolfi$^{43}$, 
C.~Adrover$^{6}$, 
A.~Affolder$^{49}$, 
Z.~Ajaltouni$^{5}$, 
J.~Albrecht$^{35}$, 
F.~Alessio$^{35}$
% ...

{\footnotesize \it
% ...
$ ^{5}$Clermont Universit\'{e}, Universit\'{e} Blaise Pascal, CNRS/IN2P3, LPC, Clermont-Ferrand, France\\
$ ^{6}$CPPM, Aix-Marseille Universit\'{e}, CNRS/IN2P3, Marseille, France\\
% ...
$ ^{11}$Physikalisches Institut, Ruprecht-Karls-Universit\"{a}t Heidelberg, Heidelberg, Germany\\
% ...
$ ^{33}$Universitat de Barcelona, Barcelona, Spain\\
$ ^{34}$Universidad de Santiago de Compostela, Santiago de Compostela, Spain\\
$ ^{35}$European Organization for Nuclear Research (CERN), Geneva, Switzerland\\
% ...
$ ^{38}$Nikhef National Institute for Subatomic Physics, Amsterdam, The Netherlands\\
% ...
$ ^{43}$H.H. Wills Physics Laboratory, University of Bristol, Bristol, United Kingdom\\
% ...
$ ^{49}$Oliver Lodge Laboratory, University of Liverpool, Liverpool, United Kingdom\\
% ...
$ ^{n}$LIFAELS, La Salle, Universitat Ramon Llull, Barcelona, Spain\\
% ...
}
\end{flushleft}

%% file: introduction.tex
\section{Introduction}

In the Standard Model (SM), the decays\footnote{Unless stated otherwise, charge conjugated modes are implicitly included throughout this paper.} \BdKstGam and \BsPhiGam proceed at leading order through the electromagnetic penguin transitions, \decay{b}{s\gamma}. At one-loop level these transitions are dominated by a virtual intermediate top quark coupling to a \W boson.
Extensions of the SM predict additional one-loop contributions that can introduce sizeable changes to the dynamics of the  transition~\cite{Descotes:theo-isospin:2011,*Gershon:th-null-tests:2006,*Mahmoudi:th-msugra:2006,*Altmannshofer:2011gn}.

Radiative decays of the \Bd meson were first observed by the CLEO collaboration in 1993 in the decay mode \mbox{\BdKstGam}~\cite{cleo:exp-first-penguins:1993}.
In 2007 the Belle collaboration reported the first observation of the analogous decay in the \Bs sector, \mbox{\BsPhiGam~\cite{belle:exp-bs2phigamma-bs2gammagamma:2007}.} 
The current world averages of the branching fractions of \mbox{\BdKstGam and \BsPhiGam} are \mbox{$(4.33\pm 0.15)\times10^{-5}$ and $(5.7^{+2.1}_{-1.8})\times10^{-5}$}, respectively~\cite{hfag:2012,babar:exp-b2kstgamma:2009,*belle:exp-b2kstgamma:2004,*cleo:exp-excl-radiative-decays:1999}. These results are in agreement with the latest theoretical predictions from NNLO calculations using soft-collinear effective theory~\cite{Ali:th-b2vgamma-NNLO:2008}, \mbox{$\BR(\BdKstGam)=(4.3\pm1.4)\times10^{-5}$ and $\BR(\BsPhiGam)=(4.3\pm1.4)\times10^{-5}$}, which suffer from large uncertainties from hadronic form factors. A better-predicted quantity is the ratio of branching fractions, as it benefits from partial cancellations of theoretical uncertainties. The two branching fraction measurements lead to a ratio  \mbox{$\BRBdKstGam/\BRBsPhiGam$=$0.7\pm0.3$}, while the SM prediction is $1.0\pm0.2$~\cite{Ali:th-b2vgamma-NNLO:2008}. When comparing the experimental and theoretical branching fraction for the \BsPhiGam decay, it is necessary to account for the large decay width difference in the $\Bs-\Bsb$ system. This can give rise to a correction on the theoretical branching fraction as large as $9\%$ as described in~\cite{deBruyn:BRmeasurement}.

The direct \CP asymmetry in the \mbox{\BdKstGam} decay is defined as \mbox{$\mathcal{A}_{\CP}=[\Gamma(\Bdb\to\overline{f})-\Gamma(\Bd\to f)]/
[\Gamma(\Bdb\to\overline{f})+\Gamma(\Bd\to f)]$.} The SM prediction, \mbox{$\mathcal{A}^\mathrm{SM}_{\CP}(\BdKstGam)=(-0.61\pm0.43)\%$\,\cite{Keum:th-b2kstgamma-pqcd:2005}}, is affected by a smaller theoretical uncertainty from the hadronic form factors than the branching fraction calculation.
The precision on the current experimental value, \mbox{$\mathcal{A}_{\CP}(\BdKstGam)=(-1.6\pm2.2\pm0.7)\%$\,\cite{babar:exp-b2kstgamma:2009,pdg2012}}, is statistically limited and more precise measurements would constrain contributions from beyond the SM scenarios, some of which predict that this asymmetry could be as large as \mbox{$-15\%$ \cite{Acp:MSSM,*Aoki:SUSY,*Aoki:SUSY-CP,*Kagan:NP}.}

This paper presents a measurement of \BRBdKstGam/\BRBsPhiGam using 1.0\invfb of data taken with the LHCb detector. The measured ratio and the world average value of \BRBdKstGam are then used to determine \BRBsPhiGam. This result supersedes a previous \lhcb measurement based on an integrated luminosity of 0.37\invfb of data at $\sqs=7\tev$~\cite{radPap}. A measurement of the direct \CP asymmetry of the decay \BdKstGam is also presented.

\section{The \lhcb detector and dataset}
The \lhcb detector~\cite{Alves:2008zz} is a single-arm forward spectrometer covering the \mbox{pseudorapidity} range $2<\eta <5$, designed for the study of particles containing \bquark or \cquark quarks. The detector includes a high precision tracking system consisting of a silicon-strip vertex detector surrounding the $pp$ interaction region, a large-area silicon-strip detector located upstream of a dipole magnet with a bending power of about $4{\rm\,Tm}$, and three stations of silicon-strip detectors and straw drift tubes placed downstream. The combined tracking system has a momentum resolution $\Delta p/p$ that varies from 0.4\% at 5\gevc to 0.6\% at 100\gevc, and an impact parameter (IP) resolution of 20\mum for tracks with high transverse \mbox{momentum (\pt)}. Charged hadrons are identified using two ring-imaging Cherenkov detectors (RICH). Photon, electron and hadron candidates are identified by a calorimeter system consisting of scintillating-pad and preshower detectors, an electromagnetic calorimeter and a hadronic calorimeter. Muons are identified by a system composed of alternating layers of iron and multiwire proportional chambers. The trigger consists of a hardware stage, based on information from the calorimeter and muon systems, followed by a software stage which applies a full event reconstruction.

Decay candidates are required to have triggered on the signal photon and the daughters of the vector meson. At the hardware stage, the decay candidates must have been triggered by an electromagnetic candidate with transverse energy \mbox{(\et) $>2.5\gev$.} The software stage is divided into two steps. The first one performs a partial event reconstruction and reduces the rate such that the second can perform full event reconstruction to further reduce the data rate. At the first software stage, events are selected when a charged track is reconstructed with IP \chisq$>16$. The IP \chisq is defined as the difference between the \chisq of the \pp interaction vertex (PV) fit reconstructed with and without the considered track. Furthermore, a charged track is required to have either \pt$>1.7\gevc$ for a photon with \et$>2.5\gev$ or \mbox{\pt$>1.2\gevc$} when the photon has \et$>4.2\gev$. At the second software stage, a track passing the previous criteria must form a \Kstarz or \Pphi candidate when combined with an additional track, and the invariant mass of the combination of the \mbox{\Kstarz (\Pphi)} candidate and the photon candidate that triggered the hardware stage is required to be within 1\gevcc of the world average \Bd (\Bs) mass. 
The data used for this analysis correspond to 1.0\invfb of \pp collisions collected in 2011 at the \lhc with a centre-of-mass energy of $\sqs=7\tev$.

Large samples of \BdKstGam and \BsPhiGam Monte Carlo simulated events are used to optimise the signal selection and to parametrise the invariant-mass distribution of the \B meson. Possible contamination from specific background channels has also been studied using dedicated simulated samples.
For the simulation, \pp collisions are generated using \pythia~6.4~\cite{Sjostrand:pythia:2006} with a specific \lhcb configuration~\cite{LHCb-PROC-2010-056}. Decays of hadronic particles are described by \evtgen~\cite{Lange:evtgen:2001} in which final state radiation is generated using \photos~\cite{Golonka:2005}. The interaction of the generated particles with the detector and its response are implemented using the \geant toolkit~\cite{Allison:geant4:2006,*Agostinelli:geant4:2003} as described in Ref.~\cite{Clemencic:2011}.


%% file: selection.tex
\section{Offline event selection}
The selection of \BdKstGam and \BsPhiGam decays is designed to maximise the cancellation of uncertainties in the ratio of their selection efficiencies. 

The charged tracks used to build the vector mesons are required to have \mbox{\pt$>500\mevc$}, with at least one of them having \pt$>1.2\gevc$. In addition, a requirement of IP $\chisq>25$ means that they must be incompatible with coming from any PV. The charged tracks are identified as either kaons or pions using information provided by the RICH system. This is based on the comparison between the two particle hypotheses. Kaons (pions) in the studied \decay{\B}{V\gamma} decays, where $V$ stands for the vector meson, are identified with a $\sim70\,(83)\,\%$ efficiency for a $\sim3\,(2)\,\%$ pion (kaon) contamination.

Photon candidates are required to have \et$>2.6\gev$. Neutral and charged clusters in the electromagnetic calorimeter are separated based on their compatibility with extrapolated tracks~\cite{Deschamps:exp-calo-reco:2003} while photon deposits are distinguished from \piz deposits using the shape of the showers in the electromagnetic calorimeter. 

Oppositely-charged kaon-pion (kaon-kaon) combinations are accepted as \mbox{$\Kstarz$ ($\Pphi$)} candidates if they form a good quality vertex and have an invariant mass within $\pm50\,(\pm10)\mevcc$ of the world average \mbox{$\Kstarz$ ($\Pphi$)} mass~\cite{pdg2012}.
The resulting vector meson candidate is combined with the photon candidate to make a \B candidate. The invariant-mass resolution of the selected \B candidate is $\approx$100\mevcc for the decays presented in this paper.

The \B candidates are required to have an invariant mass within 1\gevcc of the world average \B mass ~\cite{pdg2012} and to have \pt$>3\gevc$. They must also point to a PV, with IP \chisq$<9$, and the angle between the \B candidate momentum direction and the \B line of flight has to be less than 20\mrad. In addition, the vertex separation \chisq between the \B meson vertex and its related PV must be larger than 100. The distribution of the helicity angle $\theta_\mathrm{H}$, defined as the angle between the momentum of any of the daughters of the vector meson and the momentum of the \B candidate in the rest frame of the vector meson, is expected to follow a $\sin^2\theta_\mathrm{H}$ function for \decay{\B}{V\gamma}, and a $\cos^2\theta_\mathrm{H}$ function for the \decay{\B}{V\piz} background. A requirement of                      $|\cos\theta_\mathrm{H}|<0.8$ is therefore made to reduce \decay{\B}{V\piz} background, where the neutral pion is misidentified as a photon. Background coming from  partially reconstructed \mbox{\B-hadron} decays is reduced by requiring the \B vertex to be isolated: its \chisq must increase by more than two units when adding any other track in the event.


%% file: background.tex
\section{Signal and background description \label{section:background}}

\begin{figure}[htb]
\begin{center}
\includegraphics[width=18pc]{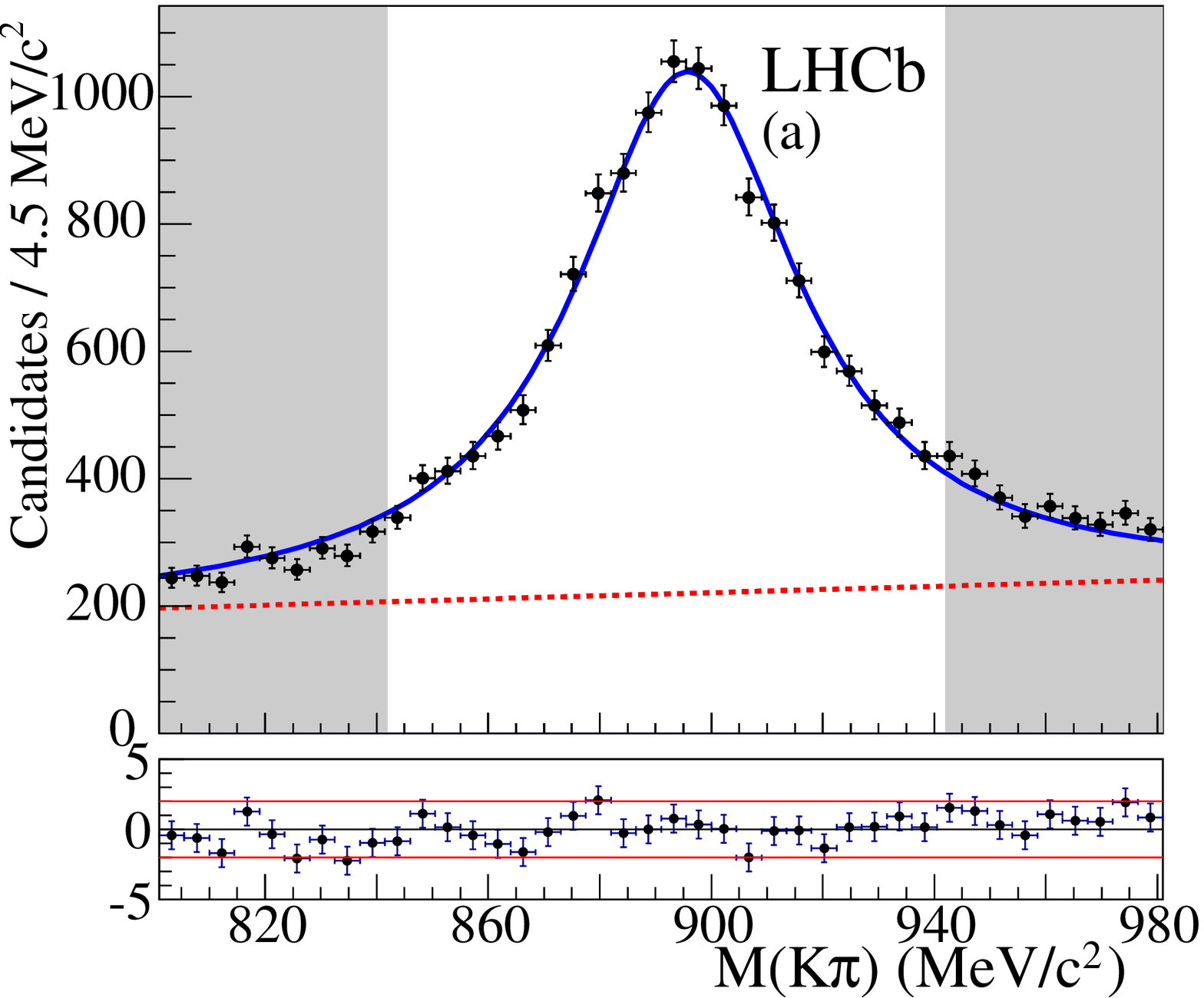} 
\includegraphics[width=18pc]{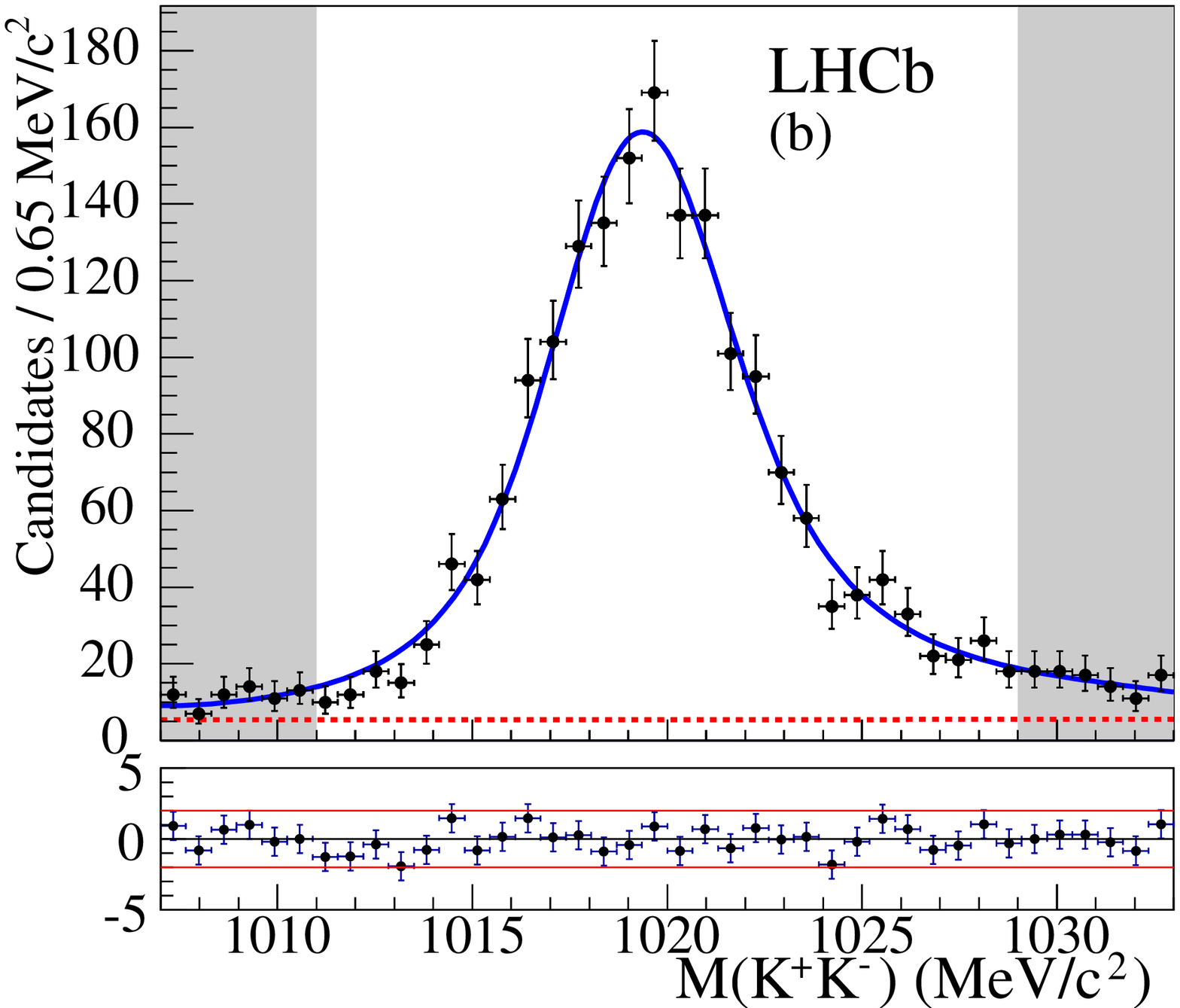}
\caption{\small Invariant-mass distributions of the (a) \Kstarz and (b) \Pphi resonance candidates. The black points represent the data and the fit result is represented as a solid blue line. The fit is described in the text. The regions outside the vector meson invariant-mass window are shaded. The Poisson \chisq residuals~\cite{Baker:chi2:1984} are shown below the fits with the $\pm2\,\sigma$ confidence-level interval delimited by solid red lines. \label{figure:vector}}
\end{center}
\end{figure}

The signal yields of the \BdKstGam and \BsPhiGam decays are determined from an extended unbinned maximum-likelihood fit performed simultaneously to the invariant-mass distributions of the \Bd and \Bs candidates. A constraint on the \Bd and \Bs masses is included in the fit which requires the difference between them to be consistent with the \lhcb measurement of \mbox{$87.3\pm0.4$\mevcc~\cite{lhcb:Bmasses}}. The \Kstarz and \Pphi resonances are described by a relativistic \mbox{$P$-wave} Breit-Wigner distribution~\cite{herab:2006} convoluted with a Gaussian distribution to take into account the detector resolution. The natural width of the resonances is fixed to the world average value~\cite{pdg2012}. A polynomial line shape is added to describe the background. The resulting distribution is fitted to the vector meson invariant-mass distribution, as shown in \fig{vector}.

The fit to the invariant mass of the vector-meson candidates yields a resonance mass of \mbox{$895.7\pm0.4$\mev} and $1019.42\pm0.09$\mev for the \Kstarz and \Pphi, respectively, in agreement with the world average values~\cite{pdg2012}. The detector resolution extracted from the fit is $5\pm4$\mev for the \Kstarz resonance and $1.3\pm0.1$\mev for the \Pphi. The effect of taking the value found in data or the world average as the central value of the vector meson mass window is negligible. In addition no systematic uncertainty due to the choice of the line shape of the resonances is assigned.

Both \BdKstGam and \BsPhiGam signal distributions are parametrised with a two-sided Crystal Ball distribution~\cite{Skwarnicki:cb:1986}. In the low mass region, there can be possible losses in the photon energy due to the fiducial volume of the calorimeter. A tail at high masses is also observed and can be explained by the spread in the error of the reconstructed \B mass and pile-up effects in the photon deposition. The parameters describing the tails on both sides are fixed to the values determined from simulation. The width of each signal peak is left as a free parameter in the fit.

The reconstructed mass distribution of the combinatorial background has been determined from the low-mass sideband of the \Kstarz mass distribution as an exponential function with different attenuation constants for the two decay channels. Additional contamination from several exclusive background decays is studied using simulated samples. The irreducible $B^0_s\to K^*\gamma$ decays, the $\Lb\to \L^*(pK^-)\gamma$ decays\footnote{$\L^*$ stands for $\L(1520)$ and other b-baryon resonances promptly decaying into a $pK^-$ final state.}, and the charmless \mbox{$B^0_{(s)}\to h^+h'^-\pi^0$} decays produce peaked contributions under the invariant-mass peak of \BdKstGam. As the experimental branching fractions of the charmless \Bs and \Lb decays are unknown, the corresponding contamination rates are estimated either using the predicted branching fraction in the case of $B_s^0\to K^{*0}\gamma$ decays, assuming SU(3) symmetry for $B^0_s\to h^+h'^-\pi^0$ decays, or by directly estimating the signal yield from an independent sample as in $\Lb\to\L^*\gamma$ decays.
The overall contribution from these decays is estimated to represent $(2.6\pm0.4)$\% and $(0.9\pm0.6)$\% of the \BdKstGam and \BsPhiGam yields, respectively. Each of these contributions is modelled with a Crystal Ball function determined from a simulated sample and their yields are fixed in the fit.

The partial reconstruction of the charged $B\to h^+h'^-\gamma X$ or $\B\to h^+h'^-\pi^0X$ decays gives a broad contribution at lower candidate masses, with a high-mass tail that extends into the signal region. The partially reconstructed $B^+\to K^{*0}\pi^+\gamma$ and $B^+\to \phi K^+\gamma$ radiative decays produce a peaking contribution in the low-mass sideband at around 5.0\gevcc for \BdKstGam and around 4.5\gevcc for \BsPhiGam. The corresponding contamination has been estimated to be $(3.3\pm1.1)\%$ and $(1.8\pm 0.3)$\% for the \BdKstGam and \BsPhiGam decays, respectively. The partially reconstructed neutral \B meson decays also contribute at the same level and several other channels exhibit a similar final state topology. These contributions are described by a Crystal Ball function and the yields are left to vary in the fit. The parameters of the Crystal Ball function are determined from the simulation. Additional contributions from the partial reconstruction of multi-body charmed decays and $B\to V\pi^0\mathrm{X}$ have been added to the simultaneous fit in the same way. The shape of these contributions, again determined from the simulation, follows an ARGUS function~\cite{ARGUS} peaking around $4.0$\gevcc. The various background contributions included in the fit model are summarised in Table~\ref{table:background}.

\begin{table}[htb!]
\begin{center}
\caption{\small  Expected contributions to the \BdKstGam and \BsPhiGam yields in the $\pm$1\gevcc mass window from the exclusive background channels: radiative decays, $h^+h'^-\gamma$ (top), charmless b decays involving energetic $\pi^0$, $h^+h'^-\pi^0$ (middle) and  partially reconstructed decays (bottom). The average measurement (exp.) or theoretical (theo.) branching fraction is given where available. Each exclusive contribution above 0.1\% is included in the fit model, with a fixed shape determined from simulation. The amplitude of the partially reconstructed backgrounds is left to vary in the fit while the $h^+h'^-\gamma$ and  $h^+h'^-\pi^0$ contributions are fixed to their expected level.}\label{table:background}
\begin{tabular}{llll}
\hline
Decay & Branching fraction  &  \multicolumn{2}{c}{Relative contribution to}   \\
      & ($\times 10^6$)     &  \BdKstGam &  \BsPhiGam\Bstrut   \\
\hline
\decay{\Lb}{\L^*\gamma}\Tstrut & estimated from data & ($ 1.0\pm 0.3$)\% & ($0.4\pm 0.3$)\%      \\
\decay{\Bs}{K^{*0}\gamma}\Tstrut\Bstrut & $1.26\pm 0.31$ (theo.~\cite{Ball:QCDfac:2007}) & ($0.8\pm 0.2$)\% & $\mathcal{O}(10^{-4})$ \\
\hline
\decay{\Bd}{K^+\pi^-\pi^0}\Tstrut & $35.9^{\,+\,2.8}_{\,-\,2.4}$ (exp.~\cite{hfag:2012})         & ($0.5\pm 0.1$)\% & $\mathcal{O}(10^{-4})$ \\
\decay{\Bs}{K^+\pi^-\pi^0}\Tstrut & estimated from SU(3) symmetry & ($0.2\pm 0.2$)\% &     $\mathcal{O}(10^{-4})$ \\
\decay{\Bs}{K^+K^-\pi^0}\Tstrut\Bstrut & estimated from SU(3) symmetry & $\mathcal{O}(10^{-4})$ & ($0.5\pm 0.5$)\% \\
\hline
\decay{\Bp}{K^{*0}\pi^+\gamma}\Tstrut & $20^{\,+\,7}_{\,-\,6}$ (exp.~\cite{hfag:2012}) &   ($3.3\pm 1.1$)\% & $ < 6\times 10^{-4}$        \\
\decay{\Bd}{K^+\pi^-\pi^0\gamma} & $41\pm4$ (exp.~\cite{hfag:2012}) & $(4.5\pm1.7)$\% &          $\mathcal{O}(10^{-4})$ \\
\decay{\Bp}{\phi K^+\gamma} & $3.5\pm0.6$ (exp.~\cite{hfag:2012}) & $3\times10^{-4}$ & $(1.8\pm 0.3)$\% \\
$B\to V\pi^0\mathrm{X}$\Tstrut\Bstrut  & $\mathcal{O}(10\%)$ (exp.~\cite{hfag:2012})  &  $\mathrm{a~few}\%$ &  $\mathrm{a~few}\%$ \\
\hline
\end{tabular}
\end{center}
\end{table}

At the trigger level, the electromagnetic calorimeter calibration is different from that in the offline analysis. Therefore, the $\pm$1\gevcc mass window requirement imposed by the trigger causes a bias in the \B meson acceptance to appear near the limits of this window. The inefficiency at the edges of the mass window is modelled by including a three-parameter threshold function in the fit model
\begin{equation}
T(m_B) = \left(1-\mathrm{erf}\left( \frac{m_B-t_{\mathrm{L}}}{\sqrt{2} \mathrm{\sigma_d}}\right)\right)\times\left(1-\mathrm{erf}\left( \frac{t_{\mathrm{U}}-m_B}{\sqrt{2} \mathrm{\sigma_d}}\right)\right)\,,
\end{equation}
where erf is the Gauss error function. The parameter $\mathrm{t_L}$($\mathrm{t_U}$) represents the actual lower (upper) mass threshold and $\mathrm{\sigma_d}$ is the resolution.


%% file: extraction.tex
\section{Measurement of the ratio of branching fractions\label{section:extraction}}

The ratio of branching fractions is measured as

\begin{equation}
\frac{\BR(\BdKstGam)}{\BR(\BsPhiGam)} = \frac{N_{\BdKstGam}}{N_{\BsPhiGam}}
                                      \times \frac{\BR(\phi\to K^+K^-)}{\BR(K^{*0}\to K^+\pi^-)}
                                      \times \frac{f_s}{f_d}
                                      \times \frac{\epsilon_{\BsPhiGam}}{\epsilon_{\BdKstGam}}\,,
\label{equation:BRRatio}
\end{equation}
where $N$ are the observed yields of signal candidates, \mbox{$\mathrm{\BR(\phi\to K^+ K^-)/\BR(K^{*0}\to K^+\pi^-)}=0.735\pm 0.008$~\cite{pdg2012}} is the ratio of branching fractions of the vector mesons, \mbox{$f_s/f_d=0.267^{+0.021}_{-0.020}$~\cite{lhcb:fsfd-paper:2011}} is the ratio of the $\B^0$ and $\B_s^0$ hadronization fractions in \pp collisions at $\sqs=7\tev$ and $\epsilon_{\BsPhiGam}/\epsilon_{\BdKstGam}$ is the ratio of total reconstruction and selection efficiencies of the two decays.

\begin{figure}[htb]
\begin{center}
\includegraphics[width=0.49\textwidth]{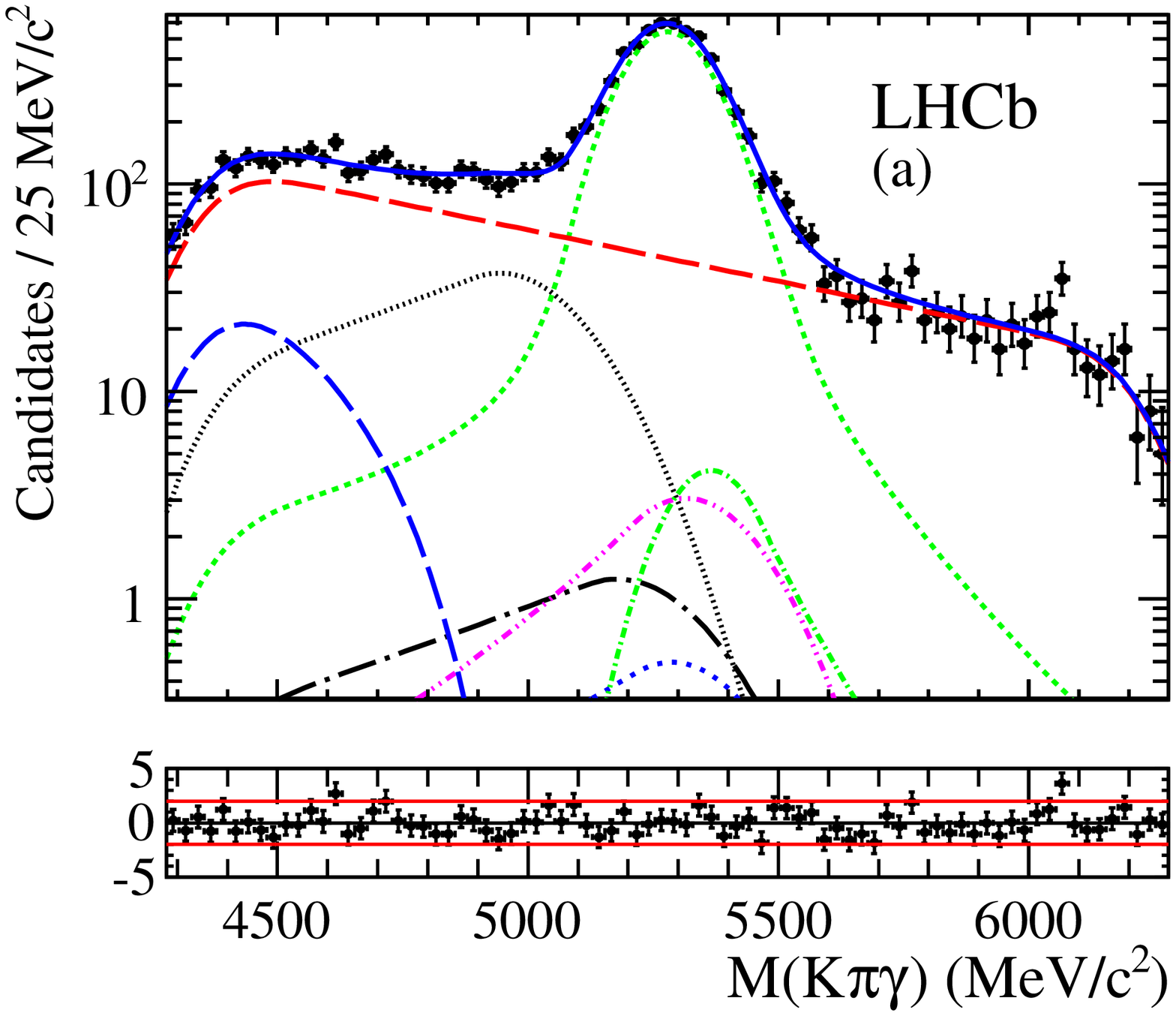} 
\includegraphics[width=0.49\textwidth]{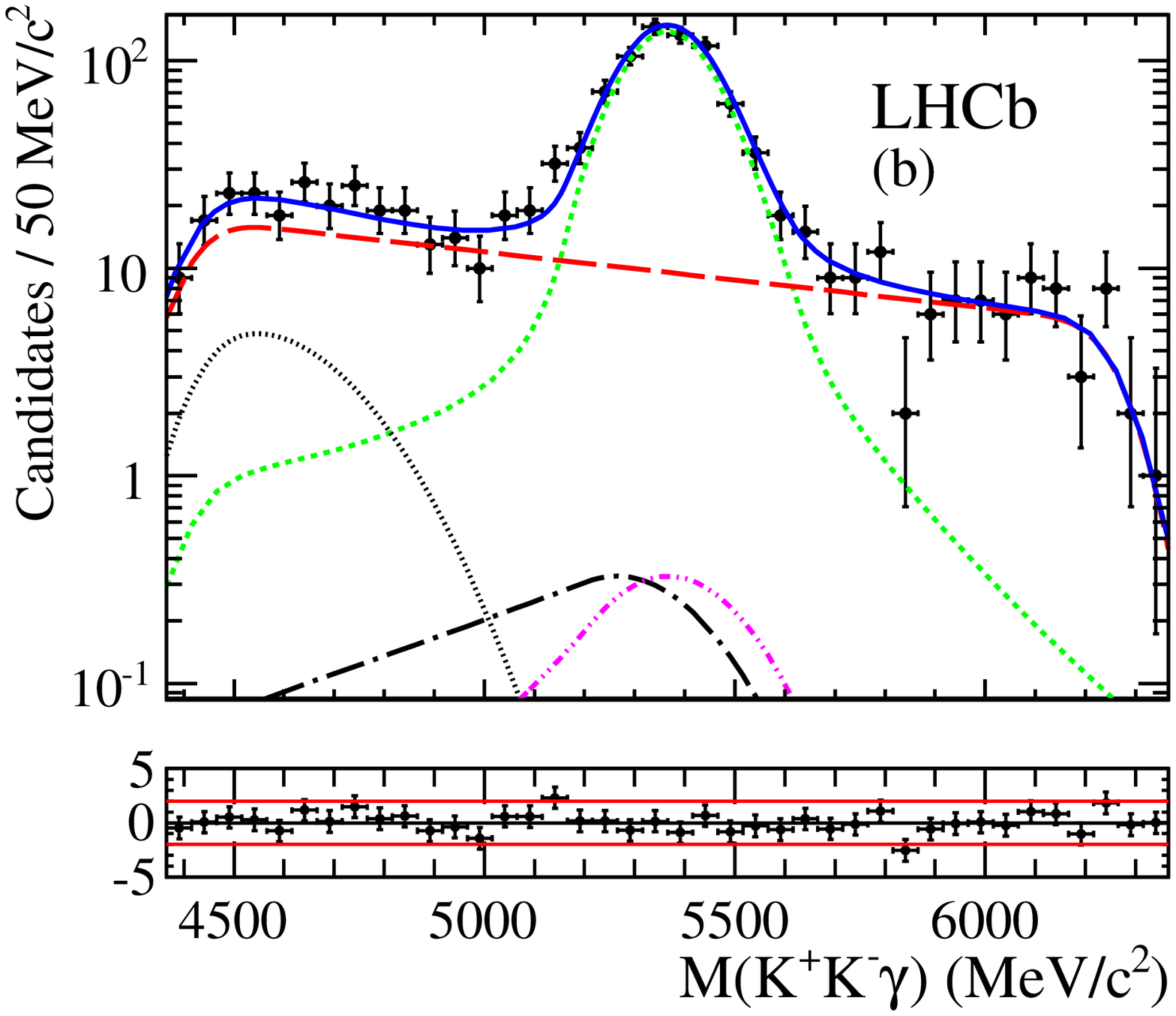}
\caption{\small Invariant-mass distributions of the (a) \BdKstGam and (b) \BsPhiGam candidates. The black points represent the data and the fit result is represented as a solid blue line. The signal is fitted with a double-sided Crystal Ball function (short-dashed green line).
The combinatorial background is modelled with an exponential function (long-dashed red line). In decreasing amplitude order, the exclusive background contributions to \BdKstGam are  \decay{B^{+(0)}}{K^{*0}\pi^{+(0)}\gamma} (short-dotted black), $B\to K^{*0}(\phi)\pi^0X$ (long-dashed blue), \decay{\Bs}{K^{*0}\gamma} (dotted short-dashed green), \decay{\Lb}{\L^*\gamma} (double-dotted dashed pink), \decay{\Bd}{K^+\pi^-\pi^0} (dotted long-dashed black) and \decay{\Bs}{K^+\pi^-\pi^0} (long-dotted blue). The background contributions to \BsPhiGam are  \mbox{\decay{B^{+(0)}}{\phi K^{+(0)}\gamma}} (dotted black), \decay{\Lb}{\L^*\gamma} (double-dotted dashed pink) and \decay{\Bs}{K^+K^-\pi^0} (dotted-dashed black). No significant contribution to \BsPhiGam is found from partially reconstructed $B\to K^{*0}(\phi)\pi^0X$ decays.
The Poisson \chisq residuals~\cite{Baker:chi2:1984} are shown below the fit with the $\pm 2\,\sigma$ confidence-level interval delimited by solid red lines. \label{figure:fits}}
\end{center}
\end{figure}

The results of the fit are shown in \fig{fits}. The number of $\BdKstGam$ and $\BsPhiGam$ candidates is $5279\pm93$ and $691\pm36$, respectively, corresponding to a yield ratio of $7.63\pm0.38$. The relative contamination from partially reconstructed radiative decays is fitted to be $(15\pm5)\%$ for \BdKstGam and $(5\pm3)\%$ for \BsPhiGam, in agreement with the expected rate from \mbox{$B^{+(0)}\to K^{*0}\pi^{+(0)}\gamma$} and \mbox{$B^{+(0)}\to \phi K^{+(0)}\gamma$}, respectively. The contribution from partial reconstruction of charmed decays at low mass is fitted to be $(5\pm4)\%$ and $(0^{+9}_{-0})$\% of the \BdKstGam and \BsPhiGam yields, respectively.

The systematic uncertainty from the background modelling is determined by varying the parameters that have been kept constant in the fit of the invariant-mass distribution within their uncertainty. The 95\% CL interval of the relative variation on the yield ratio is determined to be $[-1.2,+1.4]\%$ and is taken as a conservative estimate of the systematic uncertainty associated with the background modelling. The relative variation is dominated by the effect from the partially reconstructed background.
This procedure is repeated to evaluate the systematic uncertainty from the signal-shape modelling, by varying the parameters of the Crystal-Ball tails within their uncertainty. A relative variation of $[-1.3,+1.4]\%$ on the yield ratio is observed and added to the systematic uncertainty.
As a cross-check of the possible bias introduced on the ratio by the modelling of the mass window thresholds and the partially reconstructed background that populates the low mass region, the fit is repeated in a reduced mass window of $\pm 700$\mevcc around the world average \B meson mass. The result is found to be statistically consistent with the nominal fit. Combining these systematic effects, an overall $({}_{-1.8}^{+2.0})$\% relative uncertainty on the yield ratio is found.

The efficiency ratio can be factorised as 
\begin{equation}
\frac{\epsilon_{\BsPhiGam}}{\epsilon_{\BdKstGam}} = r_{\text{reco\&sel}}
                                               \times r_{\text{PID}}
                                               \times r_{\text{trigger}}\,,
\label{equation:ratioEffs}                                             
\end{equation}
where $r_{\text{reco\&sel}}$, $r_{\text{PID}}$ and $r_{\text{trigger}}$ are the efficiency ratios due to the reconstruction and selection requirements, the particle identification (PID) requirements and the trigger requirements, respectively.

The correlated acceptance of the kaons due to the limited phase-space in the \mbox{\decay{\Pphi}{\Kp\Km}} decay causes the \Pphi vertex to have a worse spatial resolution than the \Kstarz vertex. This affects the \BsPhiGam selection efficiency through the IP \chisq and vertex isolation cuts, while the common track cut \pt$>500\mevc$ is less efficient on the softer pion from the \Kstarz decay. These effects partially cancel and the reconstruction and selection efficiency ratio is found to be  $r_{\text{reco\&sel}}=0.906 \pm 0.007\,\text{(stat.)}\pm 0.017\,\text{(syst.)}$. The majority of the systematic uncertainties also cancel, since the kinematic selections are almost identical for both decays. The remaining systematic uncertainties include the hadron reconstruction efficiency, arising from differences in the interaction of pions and kaons with the detector and uncertainties in the description of the detector material. The reliability of the simulation in describing the $\text{IP}\,\chi^2$ of the tracks and the isolation of the \B vertex is also included in the systematic uncertainty on the $r_{\text{reco\&sel}}$ ratio. The simulated samples are weighted for each signal and background contribution to reproduce the reconstructed mass distribution seen in data. No further systematic uncertainties are associated with the use of the simulation, since kinematic properties of the decays are observed to be well modelled.
Uncertainties associated with the photon are negligible, because the reconstruction is identical in both decays.

The PID efficiency ratio is determined from data by means of a calibration procedure using pure samples of kaons and pions from \decay{\Dstarpm}{\Dz(\Kp\pim)\pipm} decays selected without PID information. This procedure yields $r_{\text{PID}}=0.839 \pm 0.005\,\text{(stat.)}\pm 0.010\,\text{(syst.)}$. 

The trigger efficiency ratio $r_{\text{trigger}}=1.080 \pm 0.009\,\text{(stat.)}$ is obtained from the simulation. The systematic uncertainty due to any difference in the efficiency of the requirements made at the trigger level is included as part of the selection uncertainty. 

Finally, the ratio of branching fractions is obtained using \eq{BRRatio},

\begin{equation*}
\frac{\BR(\BdKstGam)}{\BR(\BsPhiGam)}=1.23\pm0.06\,\mathrm{(stat.)} \pm0.04\,\mathrm{(syst.)} \pm0.10\,(f_s/f_d)\,,\label{equation:resultStat}
\end{equation*}
where the first uncertainty is statistical, the second is the experimental systematic uncertainty and the third is due to the uncertainty on $f_s/f_d$.
The contributions to the systematic uncertainty are summarised in Table \ref{table:effSummary}.

\begin{table}[htb!]
\center
\caption{\label{table:effSummary}\small Summary of the individual contributions to the relative systematic uncertainty on the ratio of branching fractions as defined in \eq{BRRatio}.}
\begin{tabular}{lc}
\hline
Uncertainty source                  & Systematic uncertainty \Tstrut\Bstrut\\ 
\hline
$r_{\mathrm{reco\& sel.}}$                      & 2.0\%\Tstrut\\
$r_{\mathrm{PID}}$                              & 1.3\% \\
$r_{\mathrm{trigger}}$   \Bstrut                & 0.8\%   \\ 
\hline
$\mathrm{\BR(\phi\to \Kp \Km)/\BR(K^{*0}\to \Kp\pim)}$\TTstrut\BBstrut  & 1.1\%\\
\hline     
Signal and background modelling  & ${}^{+2.0}_{-1.8}$\%\Tstrut\Bstrut \\
\hline
Total                                           & 3.4\%\Tstrut \\
\hline
\end{tabular} 
\end{table}


%% file: CPAsymmetry.tex

\section{Measurement of the \boldmath{\CP} asymmetry in \BdKstGam decays} \label{section:corrections}

The \BdKstGam and $\Bdb\to\Kstarzb\gamma$ invariant mass distributions are fitted simultaneously to measure a raw asymmetry defined as

\begin{equation}
\mathcal{A}_{\mathrm{RAW}}=\frac{N(\Km\pip\g)-N(\Kp\pim\g)}{N(\Km\pip\g)+N(\Kp\pim\g)}\,,
\end{equation}
where $N(X)$ is the signal yield measured in the final state $X$. This asymmetry must be corrected for detection and production effects to measure the physical \CP asymmetry. The detection asymmetry arises mainly from the kaon quark content giving a different interaction rate with the detector material depending on its charge. The \Bd and \Bdb mesons may also not be produced with the same rate in the region covered by the \lhcb detector, inducing the \Bd meson production asymmetry. The physical \CP asymmetry and these two corrections are related through

\begin{equation}
{\cal A}_{CP}(\BdKstGam) = {\cal A}_{\mathrm{RAW}}(\BdKstGam) - {\cal A}_\mathrm{D}(K\pi) -\kappa {\cal A}_\mathrm{P}(\Bd)\,,
\label{equation:CPasym}
\end{equation}
where ${\cal A}_\mathrm{D}(K\pi)$ and ${\cal A}_\mathrm{P}(\Bd)$ represent the detection asymmetry of the kaon and pion pair and \Bd meson production asymmetry, respectively. The dilution factor $\kappa$ arises from the oscillations of neutral \B mesons.

To determine the raw asymmetry, the fit keeps the same signal mean and width, as well as the same mass-window threshold parameters for the \Bd and \Bdb signal. The yields of the combinatorial background and partially reconstructed decays are allowed to vary independently. The relative amplitudes of the exclusive peaking backgrounds, $\Lb\to \L^*\gamma$, $\Bs\to K^{*0}\gamma$ and $B^0_{(s)}\to K^+\pi^-\pi^0$, are fixed to the same values for both \B flavours. 

\begin{figure}[b]
\begin{center}
\includegraphics[width=18pc]{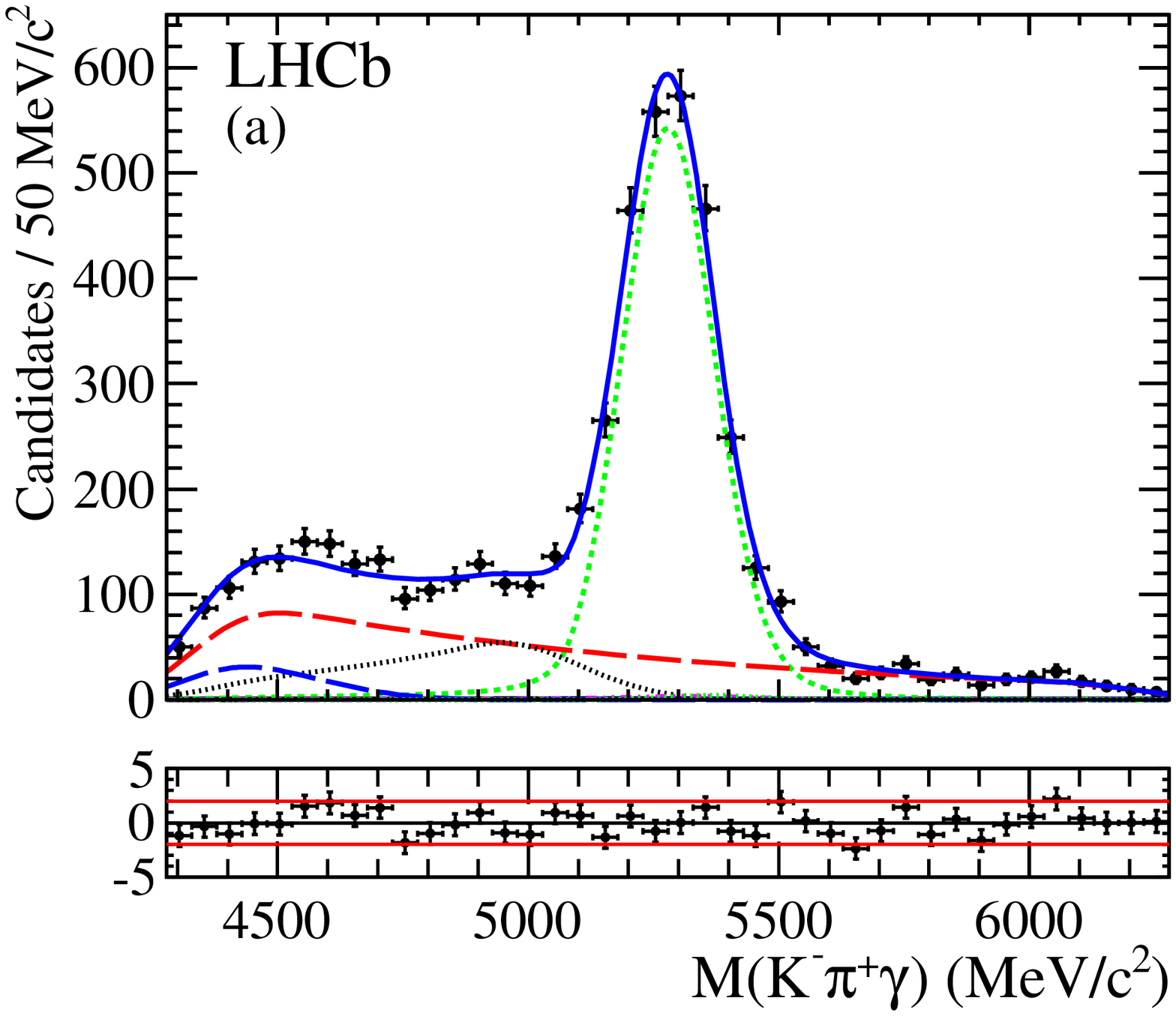}
\includegraphics[width=18pc]{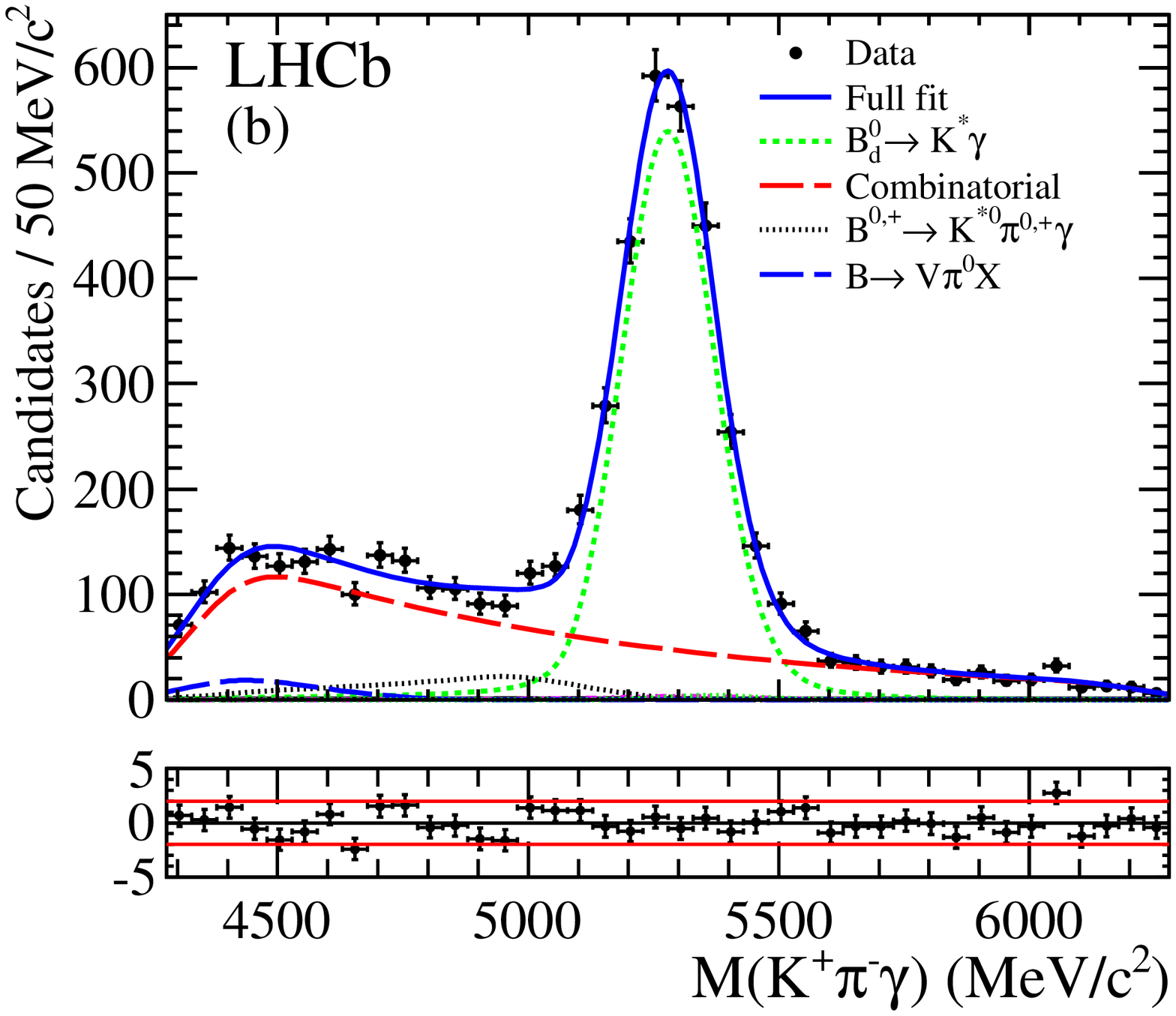}
\caption{\small Invariant-mass distributions of the (a) \Bdb$\rightarrow$\Kstarzb\g and (b) \BdKstGam decay candidates. The black points represent the data and the fit result is represented as a solid blue line. The different background components are also shown. The Poisson \chisq residuals~\cite{Baker:chi2:1984} are shown below the fits with the $\pm2\,\sigma$ confidence-level interval delimited by solid red lines. \label{figure:fitsCP}}
\end{center}
\end{figure}

\figf{fitsCP} shows the result of the simultaneous fit. The yields of the combinatorial background across the entire mass window are compatible within statistical uncertainty. The number of combinatorial background candidates is $2070\pm414$ and $1552\pm422$ in the full mass range for the \BdKstGam and $\Bdb\to\Kstarzb\gamma$ decays, respectively. The contribution from the charmless partially reconstructed decay \decay{\Bp}{\Kstarz\pip\gamma} to \BdKstGam and $\Bdb\to\Kstarzb\gamma$ is $(10\pm6)\,\%$ and $(24\pm7)\,\%$ of the signal yield, respectively. Furthermore, the charmed partially reconstructed decays $B\to \Kstarz\pi^0\mathrm{X}$ contribute with $(7\pm8)\,\%$ and $(9\pm8)\,\%$ of the signal yield to the \BdKstGam and $\Bdb\to\Kstarzb\gamma$ decays, respectively. The latter decays give contributions that are mainly located outside the signal invariant-mass region, as can be seen from \fig{fitsCP}. 

The value of the raw asymmetry determined from the fit is $\mathcal{A}_{\mathrm{RAW}} = (0.3\pm1.7)\,\%$, where the uncertainty is statistical only. 

The systematic uncertainty from the background modelling is determined as explained in \secr{background}. To address the systematic uncertainty from the possible \CP asymmetry in the background, the yield of the $\Bd\to K^+\pi^-\pi^0$ decay is varied within its measured \CP asymmetry \mbox{$\mathcal{A}_{\CP}(\Bd\to K^{*0}\pi^0)=(-15\pm 12)\%$~\cite{hfag:2012}}. For the other decays, a measurement of the \CP asymmetry has not been made. The variation is therefore performed over the full $\pm100\%$ range. The effect of these variations on $\mathcal{A}_{\mathrm{RAW}}$ gives rise to a Gaussian distribution centred at $-0.2\%$ with a standard deviation of 0.7\%, thus a correction of $\Delta {\cal A}_{\mathrm{bkg}}=(-0.2 \pm 0.7)\%$ is applied. The systematic uncertainty from the signal modelling is evaluated using a similar procedure and is found to be negligible. The possible double misidentification (\mbox{$K^-\pi^+\to\pi^- K^+$}) in the final state would induce a dilution of the measured raw asymmetry. This is evaluated using simulated events and is also found to be negligible.

An instrumental bias can be caused by the vertical magnetic field, which deflects oppositely-charged particles into different regions of the detector. Any non-uniformity of the instrumental performance could introduce a bias in the asymmetry measurement. This potential bias is experimentally reduced by regularly changing the polarity of the magnetic field during data taking. As the integrated luminosity is slightly different for the ``up" and ``down" polarities, a residual bias could remain. This bias is studied by comparing the \CP asymmetry measured separately in each of the samples collected with opposite magnet polarity, up or down. Table~\ref{table:AcpPol} summarises the \CP asymmetry and the number of signal candidates for the two magnet polarities. The asymmetries with the two different polarities are determined to be compatible within the statistical uncertainties and the luminosity-weighted average, ${\cal A}_{\mathrm{RAW}}=(0.4\pm1.7)\%$, is in good agreement with the \CP asymmetry measured in the full data sample. 

\begin{table}[ht!]
\begin{center}
\caption{\small \label{table:AcpPol} \CP asymmetry and total number of signal candidates measured for each magnet polarity.}
\begin{tabular}{lcc}
\hline
& Magnet Up & Magnet Down \Tstrut\Bstrut\\
\hline
$\int {\cal L}dt$ (\invpb) & $432\pm15$ & $588\pm21$ \Tstrut\Bstrut\\
${\cal A}^{\mathrm{RAW}}$ (\%) & $1.3\pm 2.6$ & $-0.4\pm 2.2$ \Bstrut\\ 
Signal candidates & $2189\pm65$ & $3103\pm71$ \Bstrut\\
\hline
\end{tabular}
\end{center}
\end{table}

The residual bias can be extracted from the polarity-split asymmetry as 
\begin{equation}
\Delta{\cal A}_\mathrm{M}=\left(\frac{\mathcal{L}^{\mathrm{up}}-\mathcal{L}^{\mathrm{down}}}{\mathcal{L}^{\mathrm{up}}+\mathcal{L}^{\mathrm{down}}}\right) \left(\frac{\mathcal{A}^{\mathrm{down}}_{\mathrm{RAW}}-\mathcal{A}^{\mathrm{up}}_{\mathrm{RAW}}}{2}\right)\,,
\end{equation}
which is found to be consistent with zero $\Delta{\cal A}_\mathrm{M} = (+0.1\pm0.2)\,\%$. The raw asymmetry obtained from the fit is corrected by $\Delta {\cal A}_{\mathrm{bkg}}$ and $\Delta{\cal A}_\mathrm{M}$.

The detection asymmetry can be defined in terms of the detection efficiencies of the charge-conjugate final states by
\begin{equation}
{\cal A}_\mathrm{D}(K\pi)=\frac{\epsilon(\Km\pip)-\epsilon(\Kp\pim)}{\epsilon(\Km\pip)+\epsilon(\Kp\pim)}\,.
\end{equation}
The related asymmetries have been studied at \lhcb using control samples of charm decays \cite{Carbone:exp-analysis-charmless-2body:2011}. It has been found that for \mbox{$K\pi$} pairs in the kinematic range relevant for our analysis the detection asymmetry is ${\cal A}_\mathrm{D}(K\pi)=(-1.0\pm 0.2)\%$. 

The \B production asymmetry is defined in terms of the different production rates
\begin{equation}
{\cal A}_\mathrm{P}(B^0)=\frac{R(\Bdb)-R(\Bd)}{R(\Bdb)+R(\Bd)}
\end{equation}
and has been measured at \lhcb to be ${\cal A}_\mathrm{P}(B^0)=(1.0\pm 1.3)\%$ using large samples of \mbox{\BdToJPsiKst} decays~\cite{Carbone:exp-analysis-charmless-2body:2011}. The contribution of the production asymmetry to the measured \CP asymmetry is diluted by a factor $\kappa$, defined as

\begin{equation}
\kappa=\frac{\int^{\infty}_0\cos(\Delta m_d t)e^{-\Gamma_d t}\epsilon(t) dt}{\int^{\infty}_0\cosh(\frac{\Delta\Gamma_dt}{2})e^{-\Gamma_d t}\epsilon(t) dt}\,,
\end{equation}
where $\Delta m_d$ and $\Delta\Gamma_d$ are the mass difference and the decay width difference between the mass eigenstates of the $\Bd-\Bdb$ system, $\Gamma_d$ is the average of their decay widths and $\epsilon(t)$ is the decay-time acceptance function of the signal selection. The latter has been determined from data using the decay-time distribution of background-subtracted signal candidates, the known \Bd lifetime and assuming $\Delta\Gamma_d=0$. The dilution factor is found to be $\kappa=0.41\pm 0.04$, where the uncertainty comes from knowledge of the acceptance function parameters as well as $\Gamma_d$ and $\Delta m_d$. 

\begin{table}[th!]
\begin{center}
\caption{\small Corrections to the raw asymmetry and corresponding systematic uncertainties.\label{table:breakdown}}
\begin{tabular}{lll}
\hline
Correction to $A_{RAW}$    &  & Value [\%]\Tstrut\Bstrut \\
\hline
Background model &$\Delta {\cal A}_{bkg}$\Tstrut             & $-0.2\pm0.7$     \\ 
Magnet polarity  &$\Delta {\cal A}_\mathcal{M}$\Tstrut       & $+0.1\pm0.3$     \\ 
Detection        &$-{\cal A}_\mathrm{D}(K\pi)$\Tstrut        & $+1.0\pm0.2$     \\ 
\Bd production   &$-\kappa {\cal A}_\mathrm{P}(\Bd)$\Tstrut  & $-0.4\pm0.5$     \\ 
\hline
Total            &                                          &  $+0.5\pm0.9$   \Tstrut\Bstrut \\ 
\hline
\end{tabular}
\end{center}
\end{table}

Adding the above corrections, which are summarised in Table \ref{table:breakdown}, to the raw asymmetry, the direct \CP asymmetry in \BdKstGam decays is measured to be
\begin{equation*}
{\cal A}_{CP}(\BdKstGam) = (0.8\pm1.7\,(\mathrm{stat.})\pm0.9\,(\mathrm{syst.}))\%\,.
\end{equation*}


%% file: conclusions.tex
\section{Results and conclusions}

Using an integrated luminosity of 1.0\invfb of \pp collision data collected by the \lhcb experiment at a centre-of-mass energy of $\sqs=7\tev$, the ratio of branching fractions between \BdKstGam and \BsPhiGam has been measured to be

\begin{equation*}
\frac{\BRBdKstGam}{\BRBsPhiGam} = 1.23 \pm 0.06\,\mathrm{(stat.)} \pm 0.04\,\mathrm{(syst.)} \pm 0.10\,(f_s/f_d)\label{equation:final-result}\,,
\end{equation*}
which is the most precise measurement to date and is in good agreement with the SM prediction of $1.0\pm0.2$~\cite{Ali:th-b2vgamma-NNLO:2008}.

Using the world average value $\BR(\BdKstGam)=(4.33\pm 0.15)~\times10^{-5}$~\cite{hfag:2012}, the \BsPhiGam branching fraction is determined to be
\begin{equation*}
\BRBsPhiGam = (3.5\pm0.4)\times 10^{-5}\,,
\end{equation*}
in agreement with the previous measurement~\cite{belle:exp-bs2phigamma-bs2gammagamma:2007}.
This is the most precise measurement to date and is consistent with, but supersedes, a previous \lhcb result using an integrated luminosity of 0.37\invfb~\cite{radPap}.

The direct \CP asymmetry in \BdKstGam decays has also been measured with the same data sample and found to be
\begin{equation*}
{\cal A}_{\CP}(\BdKstGam)=(0.8\pm1.7\,\mathrm{(stat.)}\pm0.9\,\mathrm{(syst.)})\%\,,
\end{equation*}
in agreement with the SM expectation of $(-0.61\pm0.43)$\,\%\,\cite{Keum:th-b2kstgamma-pqcd:2005}.
This is consistent with previous measurements~\cite{babar:exp-b2kstgamma:2009,*belle:exp-b2kstgamma:2004,*cleo:exp-excl-radiative-decays:1999}, and is the most precise result of the direct \CP asymmetry in \BdKstGam decays to date.


%% file: aknowledgments.tex
\section*{Acknowledgements}

\noindent We express our gratitude to our colleagues in the CERN accelerator
departments for the excellent performance of the LHC. We thank the
technical and administrative staff at CERN and at the LHCb institutes,
and acknowledge support from the National Agencies: CAPES, CNPq,
FAPERJ and FINEP (Brazil); CERN; NSFC (China); CNRS/IN2P3 (France);
BMBF, DFG, HGF and MPG (Germany); SFI (Ireland); INFN (Italy); FOM and
NWO (The Netherlands); SCSR (Poland); ANCS (Romania); MinES of Russia and
Rosatom (Russia); MICINN, XuntaGal and GENCAT (Spain); SNSF and SER
(Switzerland); NAS Ukraine (Ukraine); STFC (United Kingdom); NSF
(USA). We also acknowledge the support received from the ERC under FP7
and the Region Auvergne.